\documentclass[showpacs,preprintnumbers,amsmath,amssymb,superscriptaddress]{revtex4}

\usepackage{graphicx}
\usepackage{dcolumn}
\usepackage{bm}

\newcommand{\bra}[1]{\left(#1\right)}

\begin{document}

\preprint{Phys. Rev. E \textbf{77}, 016110 (2008)}

\title{Transient dynamics for sequence processing neural networks: effect of degree distributions}

\author{Yong Chen}
\altaffiliation{Email address: \tt {ychen@lzu.edu.cn}}
\affiliation{Research Center for Science, Xi'an Jiaotong University, Xi'an 710049, China}
\affiliation{Institute of Theoretical Physics, Lanzhou University, Lanzhou 730000, China}

\author{Pan Zhang}
\altaffiliation{Email address: \tt {july.lzu@gmail.com}}
\affiliation{Institute of Theoretical Physics, Lanzhou University, Lanzhou 730000, China}

\author{Lianchun Yu}
\affiliation{Institute of Theoretical Physics, Lanzhou University, Lanzhou 730000, China}

\author{Shengli Zhang}
\affiliation{Research Center for Science, Xi'an Jiaotong University, Xi'an 710049, China}

\date{\today}

\begin{abstract}
We derive an analytic evolution equation for overlap parameters, including the effect of degree distribution on the transient dynamics of sequence processing neural networks. In the special case of globally coupled networks, the precisely retrieved critical loading ratio $\alpha _c = N ^{-1/2}$ is obtained, where $N$ is the network size. In the presence of random networks, our theoretical predictions agree quantitatively with the numerical experiments for delta, binomial, and power-law degree distributions.
\end{abstract}


\maketitle

\section{Introduction}

Recently, structure and dynamics in complex networks have attracted considerable attention and have been investigated in a large variety of research fields \cite{boccaletti2006}. In particular, an important topic is whether the structure of neural wiring is related to brain functions.

Starting from pioneering milestone works that modeled Ising spin for neural networks, a large body of research has made a significant contribution to our understanding of parallel information processing in nervous tissue \cite{hopfield82,amit85,amit85-2}. The equilibrium properties of the Hopfield model in a fully connected topology with the typical Hebbian prescription for the interaction strengths have been successfully described by a replica method \cite{amit85,amit85-2}. The dynamics of the fully connected Hopfield model with static patterns and sequence patterns have been widely studied using generating functional analysis \cite{coolen,during98} and signal-to-noise analysis \cite{amari88,okada95,bolle04}.

In recent years, there have been a large number of numerical studies of the Hopfield model on the complex structure, focusing on how the topology of a network, the degree distribution in particular, affects the computational performance of the formation of associative memories \cite{mcgraw03,fontoura03,stauffer,kim04,torres}. Various random diluted models have been studied, including the extreme diluted model \cite{Derrida87,Watkin91}, the finite diluted model \cite{z07,Theumann03}, and the finite connection model \cite{WC03}. However, to the best of our knowledge, here we derive for the first time the equation of retrieval dynamics for sequence processing neural networks with complex network topology.

In this paper, we study a modification of the Hopfield network, known as the sequence processing model, which acts as a temporary associative memory model \cite{somplinsky86,amari,bauer90,coolen92}. This model is very important to understand how the nervous system allows the learning of behavioral sequences because it requires hundreds of transitions that need to be precisely stored in neuronal connections \cite{MGMTM04}. The asymmetry of the interaction matrix rules out equilibrium statistical mechanical methods of analysis, including conventional replica theory. The goal of this work is to study the effect of the degree distribution on the transient dynamics of the sequence processing neural network. Using a probability approach, we derive an analytic time evolution equation for the overlap parameter with an arbitrary degree distribution that is consistent with our extensive numerical simulation results.

This paper is organized as follows. In Section \ref{sect2} we introduce the definition of the sequence processing neural networks. The time evolution equations of the order parameters for the effects of degree distribution are derived and discussed in Section \ref{sect3}. Section \ref{sect4} contains the comparison of theoretical results with numerical simulations. Finally, Section \ref{sect5} presents a summary and the concluding remarks.

\section{Model definition}\label{sect2}

In this paper, we consider a general version of sequence processing neural networks with parallel dynamics. The model consists of $N$ Ising spin neurons $s_i \in \{-1,1\}$. If the neuron $i$ is at exciting status we put $s_i = 1$; otherwise (neuron $i$ is inhibiting), we put $s_i =-1$. The embedded patterns are $p$ states of the systems $\xi _i^{\mu} \in \{1,-1\} (\mu =1,2,\ldots,p; i=1,2,\ldots,N)$. The patterns are random so that each $\xi _i^{\mu}$ takes the values $\pm 1$ with equal probability. The couplings between neurons are represented by the following form,
\begin{equation}
J_{ij} = \frac{1}{N} \sum_{\mu=1}^p \xi_i^{\mu+\Delta p} w_{ij} \xi_j^\mu \qquad (\mu: \mathrm{mod} \ p),
\label{eq01}
\end{equation}
where $w_{ij} \in \{0,1\}$ is a matrix element to tune the connection topology of coupling matrix $\mathbf J$ and $\Delta p = 0,1,2,\ldots,N-1$ describes the patterns learned as dynamic objects. This definition is clearly a typical asymmetric neural network. In the special case of $w_{ij} =1$ and $\Delta p =0$, the synapses are symmetric as in the Hopfield-Hebb networks.

The evolution dynamics of the systems are restricted to deterministic parallel dynamics where the spins are updated simultaneously according to
\begin{equation}
s_i\left(t+1\right) = \mathrm{sgn} \left(\sum_jJ_{ij}s_j \left(t\right) \right),
\label{eq02}
\end{equation}
where $\mathrm{sgn} (x)$ is the sign function. The time step is set to $1$ in all our work.

In order to analyze the retrieval dynamics, the macroscopic overlap order parameter at time $t$ is defined by
\begin{equation}
m ^\mu (t)=\frac{1}{N}\sum_{i=1}^N \xi_i^\mu s_i(t) =\frac{2\lambda ^\mu (t)}{N} -1, \qquad \mu =1,2,\ldots,p.
\label{eq03}
\end{equation}

\begin{figure}
\includegraphics[width=0.5\textwidth]{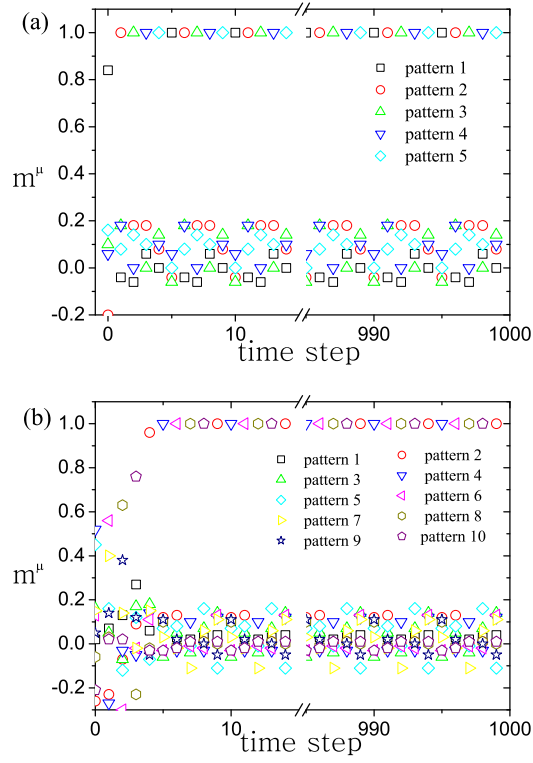}
\caption{(Color online) Temporal evolution of the macroscopic overlap parameters $m^{\mu}$ with respect to time $t$ in the case of $w_{ij} =1$. (a) $N=100$, $p=5$, and $\Delta p=1$; (b) $N=200$, $p=10$, and $\Delta p=2$.}
\label{fig1}
\end{figure}

Here, $\lambda ^\mu (t)$ is the number of spins which have the same sign between the network state at time $t$ and the given pattern ${\mathbf \xi} ^\mu$. In Fig. \ref{fig1}, we present the evolutionary processes of the order parameter $m^\mu$ for the typical sequence processing model with a global connection. It is found that the stored memories are reconstructed as a period-$(p/\Delta p)$ cycle: $2\rightarrow 3\rightarrow 4\rightarrow 5\rightarrow 1\rightarrow2 \rightarrow \cdots$ in Fig. \ref{fig1}(a), and $4\rightarrow 6\rightarrow 8\rightarrow 10\rightarrow 2 \rightarrow 4 \rightarrow \cdots$ in Fig. \ref{fig1}(b). To simplify the recent study, $\Delta p =1$ in the following context. Note that the evolution equation for overlap parameters obtained with $\Delta p=1$ is also suitable for any other $\Delta p$ values.

\section{Transient dynamics and macroscopic observable}\label{sect3}

Considering Eq. (\ref{eq01}) and Eq. (\ref{eq02}), one can get the following one-step update process,
\begin{eqnarray}
s_i\bra{t+1} &=& \mathrm{sgn} \bra{\sum_{j=1}^N \bra{ \frac{1}{N} \sum_{\mu=1}^p \xi_i ^{\mu+1} w_{ij} \xi_j^\mu}s_j\bra{t}} = \mathrm{sgn} \left( \frac{1}{N} \sum_{\mu=1}^p \xi_i^{\mu+1} \sum_{j=1}^N w_{ij} \xi_j ^\mu s_j \bra{t} \right) \nonumber \\
&=& \mathrm{sgn} \left( \frac{1}{N} \xi_i^{\nu+1} \sum_j w_{ij} \xi_j^\nu s_j \bra{t} + \frac{1}{N} \sum^p_{\mu\neq \nu} \xi_i^{\mu+1} \sum_j w_{ij} \xi_j^\mu s_j \bra{t} \right),
\label{eq04}
\end{eqnarray}
where $\mathbf \xi ^\nu$ is the $\nu$th stored pattern that is closest to state $\mathbf s(t)$ or $m ^\nu \bra{t} = \mathrm {max} \{ m^1,m^2,\ldots,m^p \}$. The contributions for a single step evolutionary process in Eq. (\ref{eq04}) consists of two parts denoted by
\begin{eqnarray}
h_i^1 \bra{t} &=& \frac{1}{N} \xi_i^{\nu+1} \sum_j w_{ij} \xi_j^\nu s_j \bra{t}, \label{eq05}\\
h_i^2 \bra{t} &=&  \sum^p_{\mu=1,\mu\neq \nu} \xi_i^{\mu+1} \bra {\frac{1}{N} \sum_j w_{ij} \xi_j^\mu s_j \bra{t} }.
\label{eq06}
\end{eqnarray}
The first part in the update function of Eq. (\ref{eq04}), $h_i^1(t)$, drives the status of the $i$th spin to $\xi_i^{\nu +1}$ at time $t+1$. The other part, $h_i^2(t)$, is the noise term. In the case of absolutely stable and precise retrieving storage, $s_i \bra{t+1} = \xi_i^{\nu +1}$. Since $s_i\bra{t+1} = \mathrm{sgn} \left( h_i^1(t) + h_i^2(t) \right)$ and $h_i^2$ cannot change the sign of $h_i^1$. If $\xi_i^{\nu+1} = +1$, we have $h_i^1>0$. To ensure that $s_i \bra{t+1} = \xi_i^{\nu +1}$, $h_i^2$ should satisfy $h_i^2 > -h_i^1$, so $ h_i^2 / h_i^1 >-1 $. If $\xi_i^{\nu+1}=-1$, $h_i^1<0$, we also have $ h_i^2 / h_i^1 >-1 $. So the probability of $s_i(t+1) = \xi_i^{\nu +1}$ is represented by the following equation,
\begin{equation}
P\bra{s_i(t+1) = \xi_i^{\nu +1}} = \sum_{z_i(t) = {-1}}^{(p-1)/m^\nu} P\bra{z_i(t)}, \quad z_i(t)= \frac{h_i^2 (t)}{h_i^1 (t)},
\label{eq07}
\end{equation}
in which $P(z_i(t))$ is the probability of $z_i(t)= h_i^2 (t)/h_i^1 (t)$. Note that the degree of node $i$ is $k_i = \sum _{j=1}^N w_{ij}$, which means that there are only $k_i$ spins that are affected. For $\xi _i^{\nu +1} =1$,
\begin{eqnarray}
h_i^1\bra{t} = \frac{k_i}{N} m^\nu.
\label{eq08}
\end{eqnarray}
Applying the above equation to Eq. (\ref{eq07}), we have
\begin{eqnarray}
P\bra{s_i(t+1) = \xi_i^{\nu +1}} &=& \sum_{h_i^2 = {-m^\nu \frac{k_i}{N} }}^{(p-1) \frac{k_i}{N} } P\bra{h_i^2(t)}, \qquad (\xi_i^{\nu +1} = 1) \label{eq09} \\
&=& \sum_{h_i^2 = {-(p-1) k_i/N}}^{m^\nu k_i/N } P\bra{h_i^2(t)}, \qquad (\xi_i^{\nu +1} = -1). \label{eq10}
\end{eqnarray}

With the following definition
\begin{eqnarray}
h_i' \bra{t} &=&  \sum^p_{\mu=1,\mu\neq \nu} \bra {\frac{1}{N} \sum_j w_{ij} \xi_j^\mu s_j \bra{t} }.
\label{eq11}
\end{eqnarray}
We find that
\begin{eqnarray}
\sum_{h_i' = {-m^\nu \frac{k_i}{N}}}^{(p-1) \frac{k_i}{N} } P\bra{h_i'(t)}
= \frac{1}{2} \bra {\sum_{h_i^2 = {-m^\nu \frac{k_i}{N} }}^{(p-1)\frac{k_i}{N}} P\bra{h_i^2} + \sum_{h_i^2 = {-(p-1) \frac{k_i}{N}}}^{m^\nu \frac{k_i}{N} } P\bra{h_i^2} }.
\label{eq12}
\end{eqnarray}
Combined with Eqs. (\ref{eq09}-\ref{eq10}), we have
\begin{eqnarray}
P\bra{s_i(t+1) = \xi_i^{\nu +1}} &=& \sum_{h_i' = {-m^\nu k_i/N }}^{(p-1)k_i/N} P\bra{h_i'(t)}.
\label{eq13}
\end{eqnarray}

Given the fact that the stored patterns are random and independent, the probability of the total number $\lambda _N^\mu$ spins that have has the same sign between $\mathbf{s}(t) $ and $\mathbf{\xi}^\mu$, $P(\lambda _N^\mu)$, is given by
\begin{equation}
P(\lambda _N^\mu)=C_N^{\lambda _N^\mu} 2^{-N}.
\label{eq14}
\end{equation}
where $C_N^{\lambda _N^\mu}$ is used to denote a binomial coefficient, i.e. $C_N^{\lambda _N^\mu} = \frac{N!}{\lambda _N^\mu !(N-\lambda _N^\mu)!}$. For the precisely retrieved case (when system can retrieve patterns without error), $\mathbf{s}(t) \approx \mathbf{\xi}^\nu$, equation (\ref{eq14}) is correct obviously. Without the precisely retrieved condition, Eq. (\ref{eq14}) in fact neglects the correlations between network states and pattern $\mu$. If network topology has a local tree structure, the above deduction is correct. If there exist many short loops in the networks, the above deriving process is just an approximation. So, in this work, our derivations below are only suitable for sparsely connected random network, where the typical loop length is about $\log_{\bar k -1} N$ and $\bar k$ is average connection. In this paper, we use the condition $N \to \infty$, $\bar k \to \infty$, and $\bar k / N \to 0$.

Then from Eq. (\ref{eq03}), we find
\begin{eqnarray}
P \bra{\frac{1}{N} \sum _j \xi _j^\mu s_j(t) = \frac{2\lambda _N^\mu}{N} -1}
= P \bra{m^\mu = \frac{2\lambda _N^\mu}{N}-1} = P \bra{\lambda _N^\mu}.
\label{eq15}
\end{eqnarray}
As noticed in the above statements [see Eqs. (\ref{eq04})-(\ref{eq06})], the local field is filtered by the topological structure of the networks. In this case, the probability of the total number $\lambda_{k_i} ^\mu$ spins that have the same sign between $\mathbf{s}(t)$ and $\mathbf{\xi}^\mu$ is
\begin{eqnarray}
P \bra{\frac{1}{N} \sum _j w_{ij} \xi _j^\mu s_j(t) = \frac{k_i}{N} m^\mu = \frac{k_i}{N} \bra{\frac{2\lambda _{k_i}^\mu}{k_i} -1}}
= P \bra{\lambda _{k_i}^\mu} = C_{k_i}^{\lambda _{k_i}^\mu} 2^{-k_i}.
\label{eq16}
\end{eqnarray}
Substituting into Eq. (\ref{eq10}) the expressions of
\begin{equation}
x_i^\mu = \frac{2\lambda _{k_i}^\mu}{N} - \frac{k_i}{N}
\label{eq17}
\end{equation}
we get the following form
\begin{eqnarray}
P \bra{\frac{1}{N} \sum _j w_{ij} \xi _j^\mu s_j(t) = x_i^\mu} & = & P \bra{\lambda _{k_i}^\mu = \frac{k_i}{2} (1+x_i^\mu)} = C_{k_i}^{\frac{k_i}{2} (1+x_i^\mu)} 2^{-k_i}.
\label{eq18}
\end{eqnarray}
The other form of $h_i'(t)$ is readily deduced from the Eq. (\ref{eq17}) which can be written as
\begin{eqnarray}
h_i'(t) &=& \sum _{\mu \neq \nu}^p x_i^\mu = \frac{2\sum _{\mu \neq \nu}^p \lambda _{k_i}^\mu}{N} - (p-1) \frac{k_i}{N}.
\label{eq19}
\end{eqnarray}
Therefore, after coarse graining like in Eqs. (\ref{eq16}, \ref{eq18}), using the definition of $\Lambda _{k_i} = \sum _{\mu \neq \nu}^p \lambda _{k_i}^\mu = \frac{Nh_i'}{2} + \frac{(p-1)k_i}{2}$, we obtain that
\begin{eqnarray}
P_1 &=& P\bra{s_i(t+1) = \xi_i^{\nu +1}} = \sum_{h_i' = {-m^\nu k_i/N }}^{(p-1)k_i/N} P\bra{h_i'(t)}
= \sum_{\Lambda _{k_i} = {\frac{(p-1)}{2} k_i - \frac{m^\nu}{2} k_i}}^{(p-1)k_i} P \bra{\Lambda _{k_i} = \sum _{\mu \neq \nu}^p \lambda _{k_i}^\mu} \nonumber \\
&=& \sum_{\Lambda _{k_i} = {\frac{(p-1)}{2} k_i - \frac{m^\nu}{2} k_i}}^{(p-1)k_i} C _{(p-1)k_i}^{\Lambda _{k_i}} 2^{-(p-1)k_i}
= \sum_{\Lambda _{k_i} = 0}^{{\frac{(p-1)}{2} k_i + \frac{m^\nu}{2} k_i}} C _{(p-1)k_i}^{\Lambda _{k_i}} 2^{-(p-1)k_i}.
\label{eq20}
\end{eqnarray}

By introducing the degree distribution of the network structure $P(k)$, where $k$ is the number of links connected to a node, the total number $\lambda ^{\nu +1}$ spins between the status $\mathbf{s}(t+1)$ and the stored pattern $\mathbf{\xi} ^{\nu +1} $ can be expressed as
\begin{eqnarray}
\lambda ^{\nu +1} = \sum _k N P_1 P(k).
\label{eq21}
\end{eqnarray}
Then from Eq. (\ref{eq03}), we obtain the macroscopic observable
\begin{equation}
m^{\nu +1} \bra{t+1} = \frac{1}{N} \sum _i \xi _i^{\nu +1} s_i(t+1)  = \sum _k 2 P_1 P\bra{k} - 1.
\label{eq22}
\end{equation}
Combining Eq. (\ref{eq20}) and Eq. (\ref{eq22}) and replacing $\Lambda _{k_i}$ by $n$, we arrive at the evolution equation of the overlap parameter
\begin{eqnarray}
m^{\nu+1}(t+1) = 2 \times \sum _k P(k) \frac{\sum_{n=0}^{\frac{1}{2}(p-1)k + \frac{1}{2}m^{\nu}(t) k} C_{(p-1)k}^n }{2^{(p-1)k}} - 1.
\label{eq23}
\end{eqnarray}

In the case of successful storage, the network finally tends to converge into a stable periodic cycle, or $m^\nu (t) = m^{\nu+1} (t + 1)$. Finally, replacing $m^\nu(t) = m^{\nu+1}(t+1)$ by $m_f$, one has the iterative solution for the final overlap parameter,
\begin{eqnarray}\label{eq24}
m_f = 2 \times \sum _k P(k) \frac{\sum_{n=0}^{\frac{1}{2}(p-1)k + \frac{1}{2}m_f k} C_{(p-1)k}^n }{2^{(p-1)k}} - 1.
\end{eqnarray}

Note that it is difficult to calculate the above iterative equation of the overlap parameter at very large $(p-1)k$ due to the computational complexity of the factorial term. A reasonable solution is to replace the binomial distribution in Eq. (\ref{eq20}) by a Gaussian distribution with the same expectation value and the same standard deviation. As a result, replacing $n$ by $x$, we reformulate Eq. (\ref{eq23}) as
\begin{eqnarray}
m^{\nu+1} (t+1) &\approx&  \sum _k 2P(k) \sum_{x=0}^{\frac{k}{2} (p-1 + m^\nu(t))} \sqrt {\frac {2}{\pi (p-1)k}} e^ {{\frac{-2 \bra{x-\frac{p-1}{2}k} ^2 }{(p-1)k}}} - 1 \nonumber \\
&=& \sum _k 2P(k) \int_{0}^{\frac{k}{2} \bra{p-1 + m^\nu(t)} } \sqrt {\frac {2}{\pi (p-1)k}} e^{{\frac{-2 \bra{x-\frac{p-1}{2}k} ^2 }{(p-1)k}}} dx - 1.
\label{eq25}
\end{eqnarray}
Substituting into the above equation the following expression,
\begin{equation}\label{eq26}
z = \frac{ x-(p-1)k/2 }{\sqrt {(p-1)k}/2},
\end{equation}
we find Eq. (\ref{eq25}) under the form
\begin{eqnarray}\label{eq27}
m^{\nu+1} (t+1) &=& \sum _k 2P(k) \sqrt{\frac{1}{2\pi}} \int_{0}^{\frac{m^\nu(t)}{\sqrt {(p-1)/k}}}  e^{-z^2/2} dz  \nonumber \\
&=& \sum _k P(k) \mathtt{erf} \left( \frac{m^\nu(t)}{\sqrt {(p-1)/k}} \right) \nonumber \\
&\approx& \sum _k P(k) \mathtt{erf} \left( \frac{m^\nu(t)}{\sqrt {p/k}} \right) ,
\end{eqnarray}
where $\mathtt {erf}(\cdot)$ is the error integral function
\begin{equation} \nonumber
\mathtt{erf}(z) = \sqrt \frac{2}{\pi} \int _0^z \exp (-x^2/2) dx.
\end{equation}
Note that in the special case of fully connected networks, $\bar k =N$ and $P(k)=\delta (k-N)$. Herein, Eq. (\ref{eq27}) is reduced to
\begin{equation}\label{eq28}
m^{\nu+1} (t+1) = \mathtt{erf} \left( \frac{m^\nu(t)}{\sqrt {p/N}} \right).
\end{equation}
which is a well-known result \cite{amari88,kinzel86}. It is interesting to compare the above equation with our result for arbitrary degree distributions. In Eq. (\ref{eq27}), there is only a little modification. It should also be mentioned that Eq. (\ref{eq27}) is also found for sparsely connected Hopfield networks \cite{zhang07}.

\section{Numerical studies}\label{sect4}

To verify the theory, we performed extensive simulations which are reported in this section. First of all, we give a succinct description of how our calculations for the iterative Eq. (\ref{eq24}) were made. Note that a precondition of our work is that the network is capable of memorizing patterns in the form of its equilibria in each trial [see Eqs. (\ref{eq07}, \ref{eq21})]. The initial overlap $m_f^0$ in the right side of the iterative equation is set to $1.0$. Then the second $m_f^1$ is obtained and is set as the initial overlap to calculate the third one. This process is repeated until $m_f^n \approx m_f^{n+1}$ within allowed precision. Thus, we arrive at the final stable macroscopic overlap parameter $m_f = m_f^n$ after $n$ iterative steps. In our simulations, it is found that the iterative procedures always converge quite rapidly, stopping after $4 - 5$ steps at most.

In order to compare the different effects of various degree distributions on the performance of neural networks, we consider the following two cases. One case is the globally coupled network. Although this case appears somewhat trivial, it is helpful to compare our study with some common conclusions. The other case is the random network with various degree distributions, including the delta function, binomial, and power-law distributions.

\subsection{Globally coupled networks}\label{sect4-1}

\begin{figure}
\includegraphics[width=0.5\textwidth]{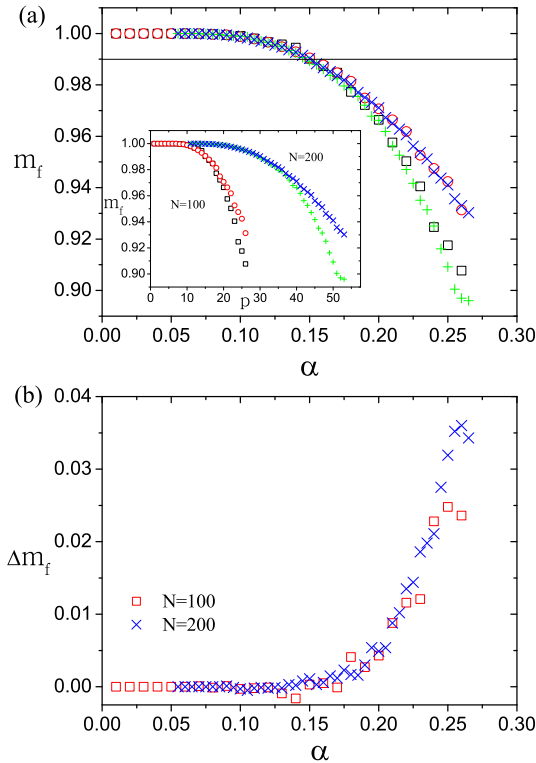}
\caption{ (Color online) The final overlap $m_f$ in the globally coupled network with $N=100, 200$. (a) ($\circ, \times$): iterative results from Eq. (\ref{eq23}), ($\square, + $): simulation results. The solid line is $m_f =0.99$ which corresponds to $\alpha \approx 0.15$ from simulations and iterative results. The inset is the same result for abscissa $p$. (b) The iterative error $\Delta m_f$ versus the loading ratio $\alpha$. Each simulation point represents an average of $200$ trials.}
\label{fig2}
\end{figure}

\begin{figure}
\includegraphics[width=0.5\textwidth]{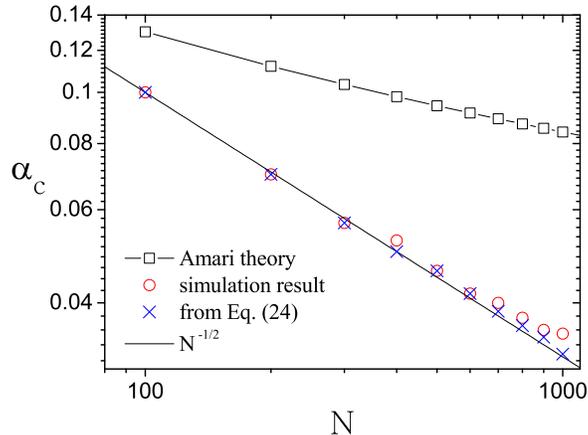}
\caption{ (Color online) The critical precise loading ratio $\alpha _c = p_c/N$ for globally coupled networks. The solid line is $\alpha _c = 1/\sqrt{N}$. ($\circ$): Simulation results. Each point represents an average of $200$ trials. ($\times$): Iterative results from Eq. (\ref{eq24}) with precision $\Delta m \leq 10^{-4}$. ($\square$): results from the Amari theory, Eq. (3.4) in \cite{amari88}.}
\label{fig3}
\end{figure}

In this case, all the neuronal spins are connected with each other at any time, namely $w_{ij} = 1$, and the degree distribution $P(k)$ is a $\delta$-function. This actually introduces a huge waste of energy but provides a neural network with the maximal retrieving performance. A large number of numerical and theoretical studies on this network dynamics have been made in the last two decades. These studies revealed that there exists a critical loading ratio known as the Amit-Gutfreund-Sompolinsky (AGS) value $\alpha_1 \approx 0.139 $ for symmetric networks \cite{amit85}, a saturated stored capacity $\alpha_s \approx 0.269 $ for sequence processing networks \cite{during98}, and the so-called exactly memorized capacity in \cite{amari88}.

It is found that the iterative form Eq. (\ref{eq24}) for the final overlap $m_f$ is effective in most successfully retrieved cases with negligible error. This is evident in Fig. \ref{fig2}, showing that $m_f$ for $N=100, 200$ until the saturation of the loading ratio $\alpha$. The iterative results are almost the same as the simulation results when $m_f$ is very close to $1$. For example, $\alpha \approx 0.15$ for $\Delta m_f =0.001$ and $\alpha \approx 0.20$ for $\Delta m_f =0.002$. When the final overlap parameters $m_f$ deviate from $1$, as the loading ratio $\alpha$ increases, the iterative error increases sharply. However, the final overlap parameter has only a small and acceptable error $\Delta m_f < 0.04$ near the saturation $\alpha_s =0.269$ [see Fig. \ref{fig2}(b)]. As stated above in Eq. (\ref{eq07}), we study only the first step behavior of the macroscopic overlap parameters. In order to get more precise results, the signal-to-noise analysis for the first few time steps should be considered \cite{bolle91,bolle04}.

One may be interested in the case where systems can retrieve stored patterns without error. We define this as the precisely retrieved case. In other words, pattern $\nu$ at time $t$ and pattern $\nu +1$ at time $t+1$ are retrieved without error, which means $m^\nu\bra{t}=1$, and $m^{\nu+1}\bra{t+1}=1$. In the stationary state, $m_f=1$. In our theory, the critical precisely retrieved storage $\alpha _c$ can be calculated from Eq. (\ref{eq24}) by setting $m_f=1$. In Fig. \ref{fig3}, we present $\alpha _c$ obtained by Eq. (\ref{eq24}) to compare with the simulation results. Obviously, $\alpha _c \approx 1/\sqrt{N}$ in the iterative algorithm Eq. (\ref{eq24}) is consistent with that in the simulation results. Here, it should be mentioned that there is another capacity definition based on absolute stability, $\alpha _A = 1/(2\log N - \log \log N)$, called the Amari capacity \cite{amari88}. Obviously, one has the following relationship between the critical loading ratios,
\begin{equation}\label{eq29}
\alpha _c < \alpha _A < \alpha _s.
\end{equation}
Note that the Amari capacity $\alpha _A$ means that there exists some probability for precise retrieval of stored patterns at least one time in a large number of trials. However, the so-called precisely retrieved capacity $\alpha _c$ in this paper means that the systems must precisely retrieve stored patterns for each trial.

\subsection{Random networks}\label{sect4-2}

\begin{figure}[h]
\includegraphics[width=0.5\textwidth]{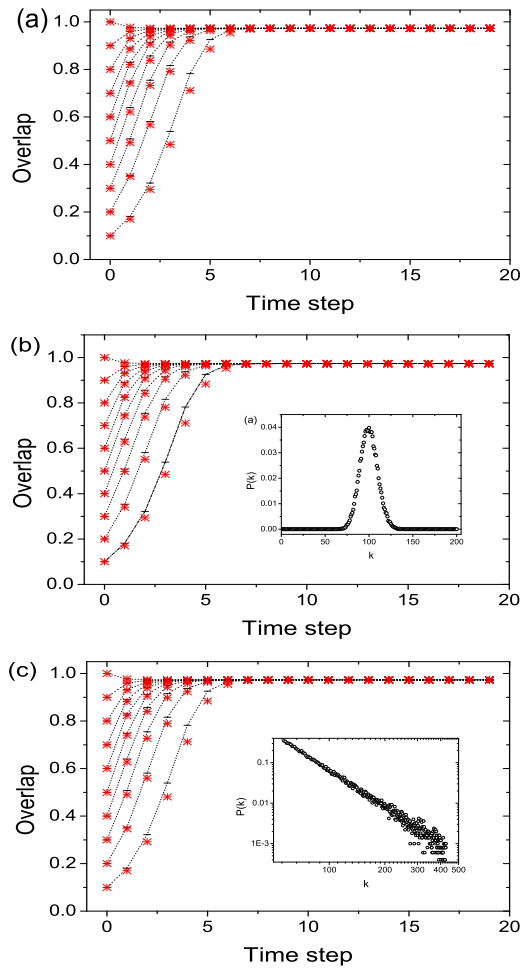}
\caption{(Color online) The temporal evolution of overlaps from our theory ($-$) and simulation ($\ast$) for (a) delta, (b) binomial, and (c) power-law degree distributions. The parameters of networks are $N=50000$, $p=20$, and the average degree $\bar{k} = 100$.}
\label{fig4}
\end{figure}

\begin{figure}
\includegraphics[width=0.5\textwidth]{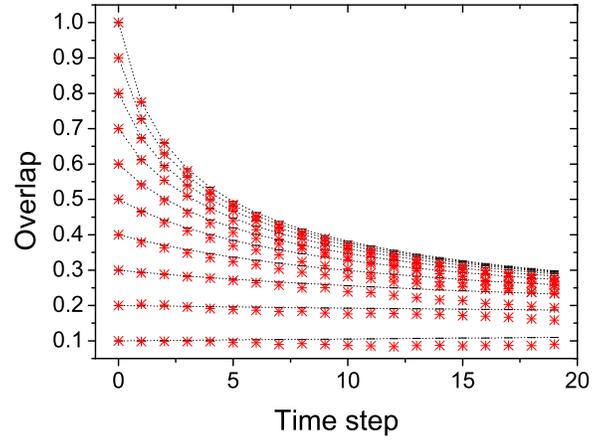}
\caption{(Color online) The temporal evolution of overlaps from our theory ($-$) and simulation ($\ast$) for power-law degree distribution $P(k) \sim k^{-3}$. The parameters of networks are $N=50000$, $p=60$, and the average degree $\bar{k} = 100$.}
\label{fig5}
\end{figure}

\begin{figure}
\includegraphics[width=0.5\textwidth]{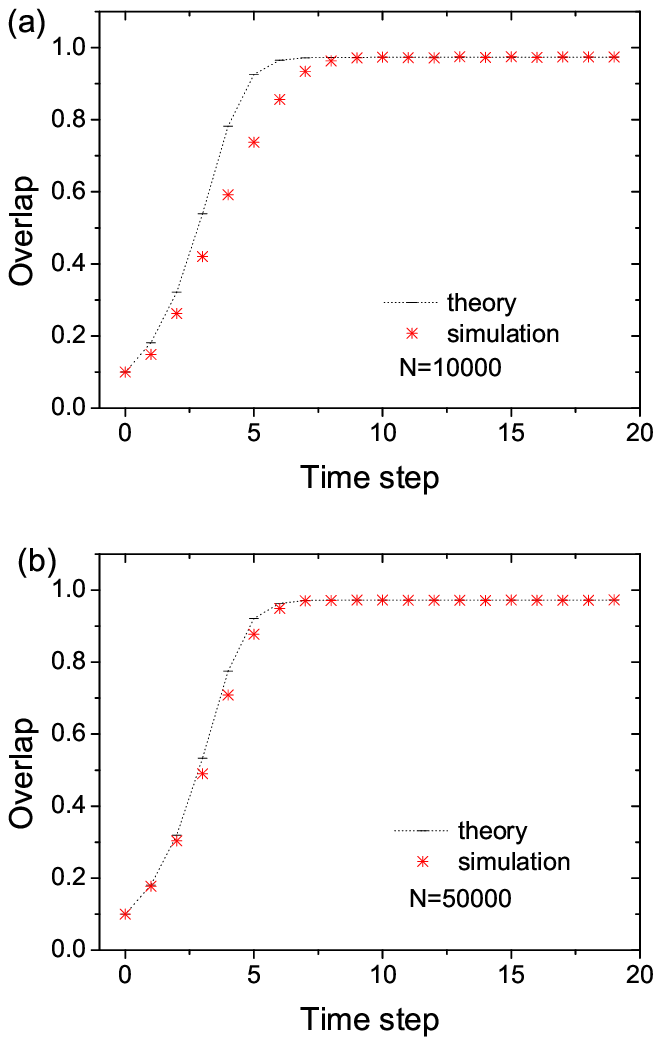}
\caption{(Color online) The temporal evolution of overlaps for binomial degree distribution with the average degree $\bar{k} = 100$, $p=20$, and (a) $N=10000$, (b) $N=50000$. }
\label{fig6}
\end{figure}

We take more general situations into consideration and explore the effect of degree distributions on the transient dynamics of neural networks in the case of random connection. In the following context, we study three situations: the delta function, binomial, and power-law degree distributions.

In Fig. \ref{fig4}, we plot the temporal evolution of overlaps for the above three types of degree distributions obtained from both the theory [Eq. (\ref{eq27})] and simulations. The parameters are: $N=50000$, the average degree $\bar k =100$, and the number of stored patterns $p=20$. The first numerical experiment is the degree distribution with the delta function
\begin{equation}
P(k)=\delta (k- \bar k).
\end{equation}
This connection topology is generated by randomizing a regular lattice whose average degree is $\bar k$. The temporal evolutions of overlap parameters from the theory and numerical simulations are plotted in Fig. \ref{fig4}(a). The second one is a binomial distribution which comes from an Erd\"{o}s-Renyi random graph \cite{renyi59} [see the inset of Fig. \ref{fig4}(b)]
\begin{equation}
P(k)=C_N^k \left( \frac{\bar k}{N} \right) ^k \left( 1- \frac{\bar k}{N} \right) ^{N-k}.
\end{equation}
The third one is the power-law distribution [see the inset of Fig. \ref{fig4}(c)],
\begin{equation}
P(k) \sim k^{-3}.
\end{equation}
The power-law degree distribution can be generated using preferential attachment \cite{albert99}. It is easy to observe that the theoretical results from our scheme are consistent with the simulations for the three degree distributions above.

Note that the presented cases are all situations of successful retrieval of the stored patterns. Fig. \ref{fig5} plots the time evolution of overlaps in the case of failed trials with $p=60$. Apparently, as stated above, encouraging results are also obtained.

Furthermore, we study the effect of size in the network on transient dynamics. Figure \ref{fig6} shows the comparison between our theory [Eq. (\ref{eq27})] and the numerical simulations for $N=10000$ and $N=50000$ in the case of binomial degree distribution. As $N$ increases under all the other same parameters, the theoretical prediction is closer to the simulation result. In fact, this size effect comes from the loop structure in networks. In this paper, our equation (\ref{eq27}) does not take into account the loop structure. Loop structure refers to the existence of (perhaps many) short loops in the network, such as triangles or quadrangles (See Ref. \cite{ZC07}). These short loops may cause the coupling of the order parameters at different times and complicate the dynamics. \cite{ZC07} suggested a parameter, loopiness coefficient, to investigate the effect of loop in the networks. Loopiness coefficients grow with $\bar{k}/N$ in random network. Our formula can present better performance for loopiness coefficients smaller with increasing sparseness of network connectivity.

\section{Summary and concluding remarks}\label{sect5}

In this paper, we have discussed the effect of degree distribution on transient dynamics for sequence processing neural networks. When the effect of loop structure is absent, we derived the analytic evolution equation for the overlap parameter [Eq. (\ref{eq27})] including the effect of degree distribution that is also obtained in the sparsely connected Hopfield model \cite{zhang07}. In the case of globally coupled networks, the so-called precisely retrieved capacity $\alpha _c = N ^{-1/2}$ is suggested by both the theory and simulations; whereas in the case of random networks, our theoretical predictions are consistent with the numerical simulation results under three situations, including the delta, binomial, and power-law degree distributions.

It should be mentioned that, in our presented work, the most efficient arrangement for storage and retrieval of patterns in sequence by the artificial neural network is the random topology. But in real brains, the topology of neural systems appears more complicated and the effect of loop structure becomes inevitable~\cite{WS98,BHO99}. In a special case, the role of loop structure has been studied without the effect of degree distribution \cite{ZC07}. In future work, we will focus on how to combine the effects from the degree distribution and the loop structure.

\section*{Acknowledgment}

This work was supported by the National Natural Science Foundation of China under Grant No. $10305005$ and by the Special Fund for Doctor Programs at Lanzhou University.

\end{document}